\documentclass[%
 reprint,
superscriptaddress,
preprintnumbers,
nofootinbib,
 amsmath,amssymb,
 aps,
]{revtex4-1}

\usepackage{graphicx}
\usepackage{dcolumn}
\usepackage{bm}
\usepackage[colorlinks=true, linkcolor=red, citecolor=blue, urlcolor=blue]{hyperref}
\usepackage{lipsum}
\usepackage{float}
\usepackage{xcolor}
\usepackage{soul}

\newcommand{\km}[1]{\textcolor{blue}{#1}}

\begin{document}


\title{Revisiting ultrahigh-energy constraints on decaying superheavy dark matter}

\author{Saikat Das}
\email{saikat.das@yukawa.kyoto-u.ac.jp}
\affiliation{Center for Gravitational Physics and Quantum Information, Yukawa Institute for Theoretical Physics, Kyoto University, Kyoto 606-8502, Japan}

\author{Kohta Murase}
\email{murase@psu.edu}
\affiliation{Department of Physics; Department of Astronomy \& Astrophysics; Center for Multimessenger Astrophysics, Institute for Gravitation and the Cosmos, The Pennsylvania State University, University Park, Pennsylvania 16802, USA}
\affiliation{School of Natural Sciences, Institute for Advanced Study, Princeton, New Jersey 08540, USA}
\affiliation{Center for Gravitational Physics and Quantum Information, Yukawa Institute for Theoretical Physics, Kyoto University, Kyoto 606-8502, Japan}

\author{Toshihiro Fujii}
\email{toshi@omu.ac.jp}
\affiliation{Graduate School of Science, Osaka Metropolitan University, Osaka 558-8585, Japan
}
\affiliation{Nambu Yoichiro Institute of Theoretical and Experimental Physics, Osaka Metropolitan University, Osaka 558-8585, Japan
}





\date{\today}
             

\begin{abstract}
We revisit constraints on decaying very heavy dark matter (VHDM) using the latest ultrahigh-energy cosmic-ray (UHECR; $E\gtrsim 10^{18}$ eV) data and ultrahigh-energy (UHE) $\gamma$-ray flux upper limits, measured by the Pierre Auger Observatory. We present updated limits on the VHDM lifetime ($\tau_\chi$) for masses up to $\sim10^{15}$~GeV, considering decay into quarks, leptons, and massive bosons. In particular, we consider not only the UHECR spectrum but their composition data that favors heavier nuclei. Such a combined analysis improves the limits at $\lesssim 10^{12}$ GeV because VHDM decay does not produce UHECR nuclei. We also show that the constraints from the UHE $\gamma$-ray upper limits are $\sim10$ more stringent than that obtained from cosmic rays, for all of the Standard Model final states we consider. The latter improves our limits to VHDM lifetime by a factor of two for dark matter mass $\gtrsim10^{12}$ GeV.
\end{abstract}

\maketitle



\section{\label{sec:intro}Introduction\protect}
The highest-energy cosmic-ray particles exhibit prominent features in their energy spectrum \cite[see, e.g.,][for recent reviews]{Anchordoqui:2018qom, AlvesBatista:2019tlv, PierreAuger:2019fdm}. The most intriguing of them is the steep decline in the flux beyond $\approx6\times10^{19}$ eV, known as the ``cutoff''. It can arise from the maximum acceleration energy inside astrophysical sources or due to interactions with the cosmic background photons during the propagation. The unprecedented amount of data collected in the last decade by the Pierre Auger Observatory (Auger) in Mendoza, Argentina, and the Telescope Array experiment (TA) in Utah, the United States \citep{PierreAuger:2020qqz, PierreAuger:2021hun, PierreAuger:2019fdm, Abbasi:2021hk, Abbasi:2021hO} has led to substantial progress in the modeling of UHECR sources and their extragalactic propagation \citep[e.g.,][]{PierreAuger:2016use, AlvesBatista:2018zui, Das:2018ymz, Heinze:2019jou, Das:2020nvx, Jiang:2020arb}. The hybrid detection technique for extensive air showers, involving fluorescence detectors in addition to the surface detectors, enables precise measurement of the energy and arrival direction of the UHECRs \citep{TelescopeArray:2018xyi, PierreAuger:2014sui}. 

However, it is difficult to determine the mass of primary UHECR nuclei in a model-independent way. Different hadronic interaction models lead to a discrete interpretation of the shower distribution data. A univocal conclusion is that the composition becomes progressively heavier with increasing energy above $\approx10^{18.2}$ eV \cite{PierreAuger:2014sui, PierreAuger:2014gko, Bellido:2017cgf, PierreAuger:2018gfc}. A variety of steady and transient astrophysical source classes, either individually or cumulatively, can account for the observed flux. Tidal disruption events, gamma-ray bursts, active galactic nuclei, starburst galaxies, and compact object mergers are some of the prominent candidates studied in the literature \citep[e.g.,][]{Meszaros:2019xej, AlvesBatista:2019tlv}.

In addition to the astrophysical sources, a cosmological interpretation is also tenable. Cosmic rays originating in the decay or annihilation of dark matter (DM), with a mass up to the grand unification energy $\Lambda_{\rm GUT} \sim10^{15}-10^{16}$ GeV may produce a non-negligible event rate beyond the cutoff in the UHECR spectrum and also within the observed energy range \cite{Medina-Tanco:1999fld, Dubrovich:2003jg, Aloisio:2007bh, Kalashev:2007ph, Aloisio:2015lva, Kalashev:2017ijd, Supanitsky:2019ayx}. While the mass of thermal relics is typically expected to be less than $\sim100$ TeV due to the unitarity bound \citep{Griest_1990, Harigaya:2016nlg}, the VHDM particles are not required to be in thermal equilibrium with the primordial plasma and have a lifetime longer than the age of the universe \citep{Chung:1998zb}. Such particles may be produced by fluctuations in the gravitational fields during the nonadiabatic expansion of the universe at early epochs, transitioning from inflationary to a matter- or radiation-dominated universe \cite{Kim:2019udq, Baker:2019ndr, Kramer:2020sbb, Ling:2021zlj}.


Current multimessenger experiments aiming to detect cosmic rays, neutrinos, and gamma rays enable us to search for the signal from the fragmentation of DM particles. The final state Standard Model (SM) particles eventually lead to $p$, $\overline{p}$, $\gamma$, $\mathrm{e^\pm}$, $\nu$, and $\overline{\nu}$, which are particle messengers used for probing the Universe. Extragalactic $\gamma$-rays reaching $\gtrsim 10^{15}$ eV are greatly attenuated, so that very-high-energy $\gamma$-rays are more powerful for studying Galactic sources \citep{HESS:2016pst, HAWC:2019tcx, LHAASO:2021cbz}. Multimessenger constraints on DM decay (or annihilation) have been studied earlier in great detail \cite{Kachelriess:2007aj, Yuksel:2007ac, Murase:2012xs, Esmaili:2012us, Murase:2015gea, Kachelriess:2018rty, Ishiwata:2019aet, Guepin_2021, Arguelles:2022nbl}.


In this work, we restrict ourselves to the mass range $10^{9} \text{\ GeV} \lesssim m_\chi \lesssim 10^{15}$ GeV and their contribution to the UHE cosmic rays and $\gamma$-rays. For $p+\overline{p}$ fluxes, we include both Galactic and extragalactic DM components, as well as an astrophysical component with a mixed composition, which improves the lower limits on $\tau_\chi$. We use the latest integrated $\gamma$-ray flux upper limits from Auger \citep{Savina:2021cva, PierreAuger:2022uwd, PierreAuger:2022aty}, for the first time in this work, and provide improved bounds on $\tau_\chi$ than obtained in earlier studies. We describe the physical scenario and the numerical methods involved in Sec.~\ref{sec:model}. We present our results in Sec.~\ref{sec:results} and discuss them in Sec.~\ref{sec:discussions}. We draw our conclusions in Sec.~\ref{sec:conclusions}.


\section{\label{sec:model}UHECRs and UHE gamma rays from VHDM decay\protect}
\begin{figure*}
\centering
\includegraphics[width=0.98\textwidth]{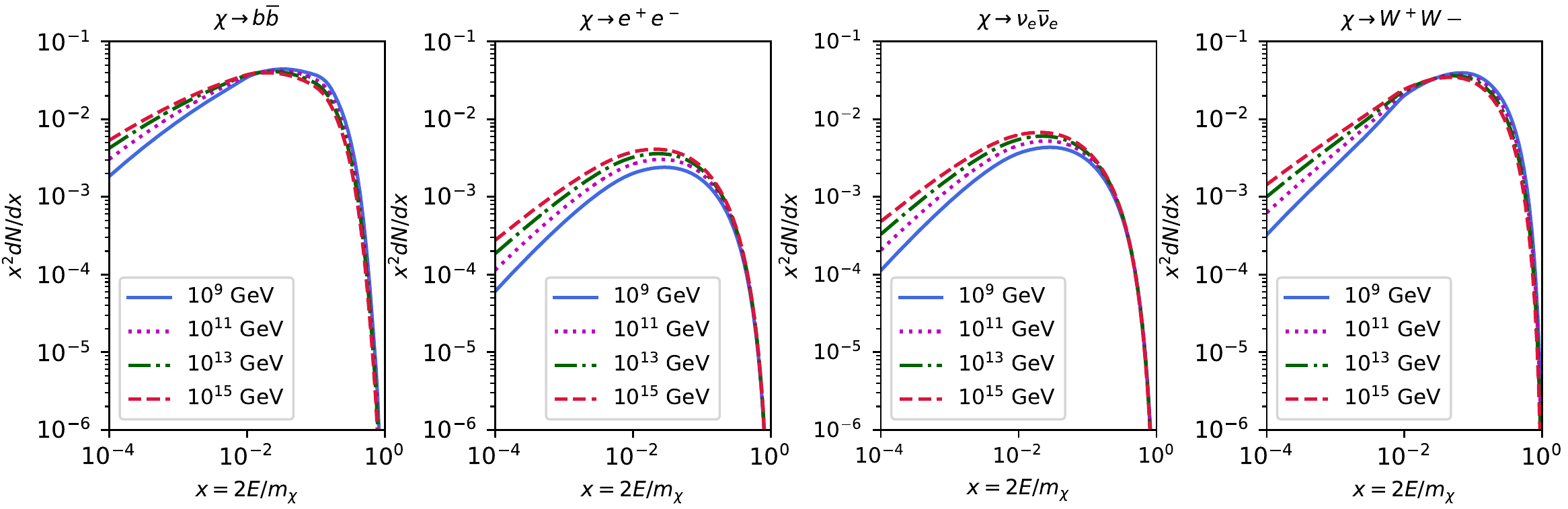}
\caption{\small{$p+\overline{p}$ prompt spectra from various DM decay modes into Standard Model particles. The different line styles represent the energies $m_\chi=10^{10}$, $10^{13}$, $10^{16}$, $10^{19}$ GeV as indicated.}}
\label{fig:prompt}
\end{figure*}
The flux of an element $S$ produced in DM decay ($\chi\rightarrow S + ...$) is given by the prompt spectrum 
\begin{align}
\dfrac{dN_s}{dx} = \dfrac{1}{\Gamma_0}\dfrac{d\Gamma}{dx},
\end{align}
where $\Gamma$ is the inclusive decay rate of $\chi$ to $S$ and $\Gamma_0=1/\tau_\chi$ is the inverse lifetime of the decay. Here $x=2E/m_\chi$ is a dimensionless variable. We assume the decay initiates through the process $\chi\rightarrow X\overline{X}$, for an arbitrary standard model particle $X$, where the particle and antiparticle each carry energy $m_\chi/2$. Finally, $X$ and $\overline{X}$ evolve to produce S, which carries a fraction $x$ of the initial energy. The local energy budget in DM (energy per unit time per unit comoving volume) is then expressed as
\begin{align}
Q_\chi^{\rm dec} = \dfrac{\rho_\chi}{m_\chi} \dfrac{m_\chi c^2}{\tau_\chi} = \dfrac{\rho_\chi c^2}{\tau_\chi}
\end{align}
because each decay event injects an energy $m_\chi c^2$, where $\rho_\chi$ is the DM density.

We use the numerical framework \textsc{HDMSpectra} as demonstrated in Ref.~\cite{Bauer:2020jay} to calculate the DM decay spectrum for energies beyond the electroweak symmetry breaking up to the GUT energy scale at $\sim10^{15}-10^{16}$ GeV. It estimates the fragmentation function $D_a^b(x; \mu_Q, \mu_0)$ to find the probability of an initial sate $a$ at an energy scale $\mu_Q$ producing a final state $b$ at $\mu_0$ that carries a momentum fraction $x$. We show the $p+\overline{p}$ fluxes in Fig.~\ref{fig:prompt} for $m_\chi=10^{15}$ GeV, corresponding to some of the widely considered decay modes in the literature.

N-body simulations of pure cold DM yield the generalized Navarro-Frenk-White (NFW) profile for DM density distribution in our Galaxy \cite{Navarro:1996gj},
\begin{align}
\rho_{\rm NFW} (R) = \dfrac{\rho_0}{(R/R_s)^\beta(1+R/R_s)^{3-\beta}}
\end{align}
where $\beta=1$ for NFW model and $R_s=11$ kpc is the scale radius. 
The Einasto profile considers the logarithmic slope to vary continuously with radius \citep{Navarro:2003ew, Einasto_1989}
\begin{align}
\rho_{\rm Ein}(R) = \rho_0\exp[-2\alpha^{-1}((R/R_{-2})^\alpha-1)] \label{eqn:einasto}
\end{align}
where $R_{-2}$ is the radius at which the logarithmic slope $d\ln\rho/d\ln R = -2$ and is given by $R_{-2}=(2-\beta)R_s$. The shape parameter is fixed to $\alpha=0.16$ \cite{Wang:2019ftp}. We take $R_h=100$ kpc as the size of the Galactic halo and $R_{\rm sc}=8.34$ kpc as the distance between the Sun and the Galactic center. The DM density in the solar neighborhood is taken as $\rho_{\rm sc} c^2=0.43$ GeV/cm$^3$ \cite{Karukes:2019jxv}, which gives the constant $\rho_0$. The boundary of the halo in the angular direction $\theta$ is
\begin{align}
s_{\rm max}(\theta) = R_{\rm sc}\cos\theta + \sqrt{R_h^2 - R_{\rm sc}^2\sin^2\theta}
\end{align}
The line-of-sight component of the flux of $S$ from direction $\theta$ is then,
\begin{align}
\Phi(E,\theta) &= \dfrac{1}{4\pi m_{\rm \chi}\tau_{\rm \chi}} \dfrac{dN_s}{dE} \int_{0}^{s_{\rm max}(\theta)}  \rho_\chi(R(s))ds \nonumber \\
&= \dfrac{\rho_{\rm sc} R_{\rm sc}}{4\pi m_\chi\tau_\chi} \dfrac{dN_s}{dE}\mathcal{J}^{\rm dec}(\theta), \label{eqn:los}
\end{align}
which yields the Galactic contribution of DM decay by performing the following integration up to $\theta=\pi$.
\begin{align}
\Phi_{\rm G}(E, \leq\theta) = \dfrac{\rho_{\rm sc} R_{\rm sc}}{4\pi m_\chi\tau_\chi} \dfrac{dN_s}{dE} \mathcal{J}^{\rm dec}_{\Omega}\\
\text{where, \ \ \ } \mathcal{J}_{\Omega}^{\rm dec} = \dfrac{2\pi}{\Omega}\int_{0}^{\theta}  \sin\theta d\theta \mathcal{J}^{\rm dec}(\theta). \label{eqn:phi_gal}
\end{align}
Here, $\Omega=2\pi(1-\cos\theta)$ is the solid angle of the field of view, and the integration in Eqn.~\ref{eqn:los} is carried out by changing the variable from line-of-sight coordinate $s$ to Galactocentric distance $R$. 

For the extragalactic case, we assume a uniform DM density distribution in the comoving distance range of 1~Mpc to 5~Gpc. We use the publicly available code \textsc{CRPropa 3} to simulate the cosmological propagation of cosmic-ray spectrum resulting from prompt DM decay \cite{AlvesBatista:2016vpy}. The cosmic-ray protons undergo various energy loss processes, viz., photomeson production, Bethe-Heitler pair creation, and $\beta$-decay of secondary neutrons. In addition, all particles lose energy due to the adiabatic expansion of the universe. The resulting flux can be expressed as
\begin{align}
\Phi_{\rm EG}(E) = \dfrac{c \Omega_\chi \rho_c}{4\pi m_\chi \tau_\chi}  \int dz \bigg|\dfrac{dt}{dz}\bigg| F(z) \int dE' \dfrac{dN'_s}{dE'}\dfrac{d\eta}{dE}(E, E',z)
\end{align}
where $dN'_s/dE'$ is the injection spectrum from prompt decay of DM and $d\eta/dE$ is the fraction of cosmic-ray protons (or antiprotons) produced with energy $E$ from parent particle of energy $E'$. The redshift evolution of cosmic-ray injection for the DM case is considered to be $F(z)=1$. Here, $\rho_c$ is the critical density in a flat FRW universe and $\rho_\chi = \Omega_\chi\rho_c$. We take $\Omega_\chi h^2=0.113$ and $\rho_cc^2h^{-2}=1.05\times10^{-5}$ GeV cm$^{-3}$, where $h$ is the dimensionless Hubble constant. The cosmological line element $|dt/dz|$ is expressed as
\begin{align}
\bigg|\dfrac{dt}{dz}\bigg| = \dfrac{1}{H_0(1+z)\sqrt{(1+z)^3\Omega_m + \Omega_\Lambda}}
\end{align}
where we take the Hubble constant at the present epoch as $H_0=67.3$ km s$^{-1}$ Mpc$^{-1}$, and the matter density as $\Omega_m = 0.315$ and the vacuum energy density $\Omega_\Lambda=0.685$ \citep{Planck:2018vyg}.
Finally, the total cosmic-ray flux from VHDM decay is obtained as $\Phi_\chi(E) = \Phi_{\rm G}(E) + \Phi_{\rm EG}(E)$. We consider a null magnetic field for the propagation of extragalactic cosmic rays because the diffusion effects are practically negligible at such high energies.

However, for calculations involving $\gamma$-ray constraints on $\tau_\chi$, we consider the Galactic DM only. The extragalactic $\gamma$-ray flux is severely attenuated by interactions in the extragalactic background light (EBL) by virtue of electromagnetic cascades and, as a result, contributes at an energy below $\approx 10^{15}$ eV. We analyze several DM decay modes of $X\overline{X}$ type, consisting of quarks, leptons, and bosons. Subsequently, we add an astrophysical component to the DM decay CR flux. For astrophysical UHECR injection, the redshift evolution $F(z)$ can be a generic power law in redshift or depend on the specific source class. The best-fit source parameters of the astrophysical component depend on the energy range of the combined spectrum and composition fit.


\section{\label{sec:results}Results \protect}
\begin{figure*}
\centering
\includegraphics[width = 0.49\textwidth]{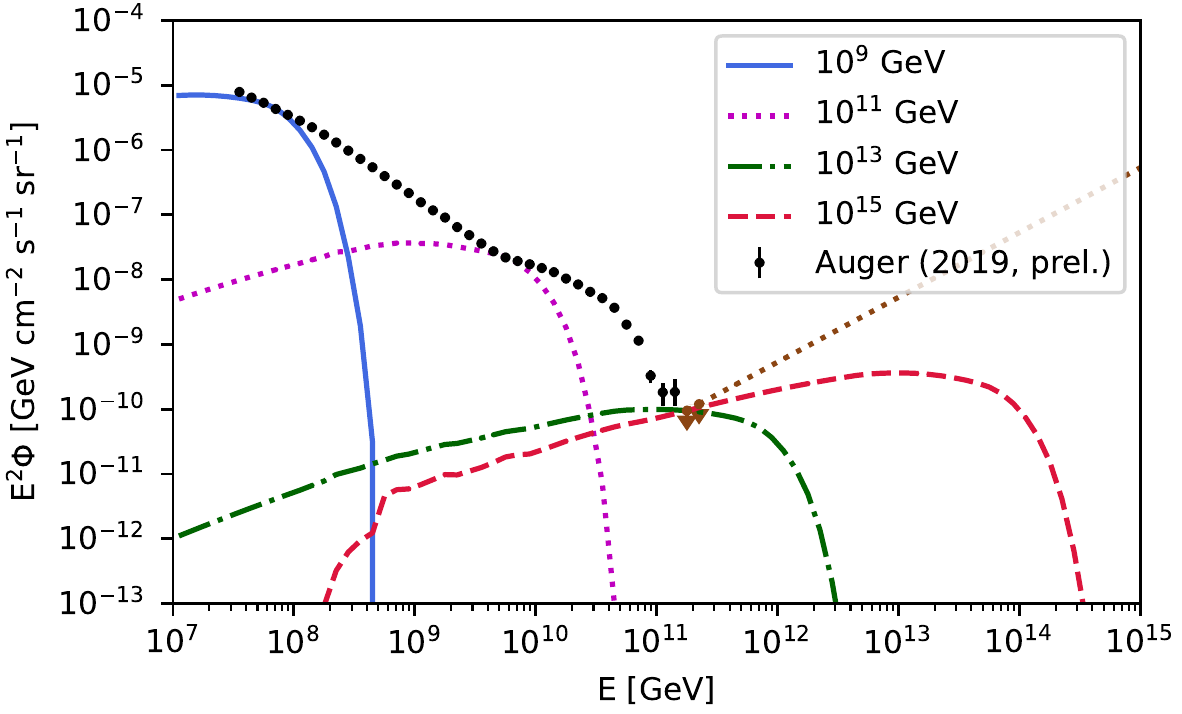}
\includegraphics[width = 0.49\textwidth]{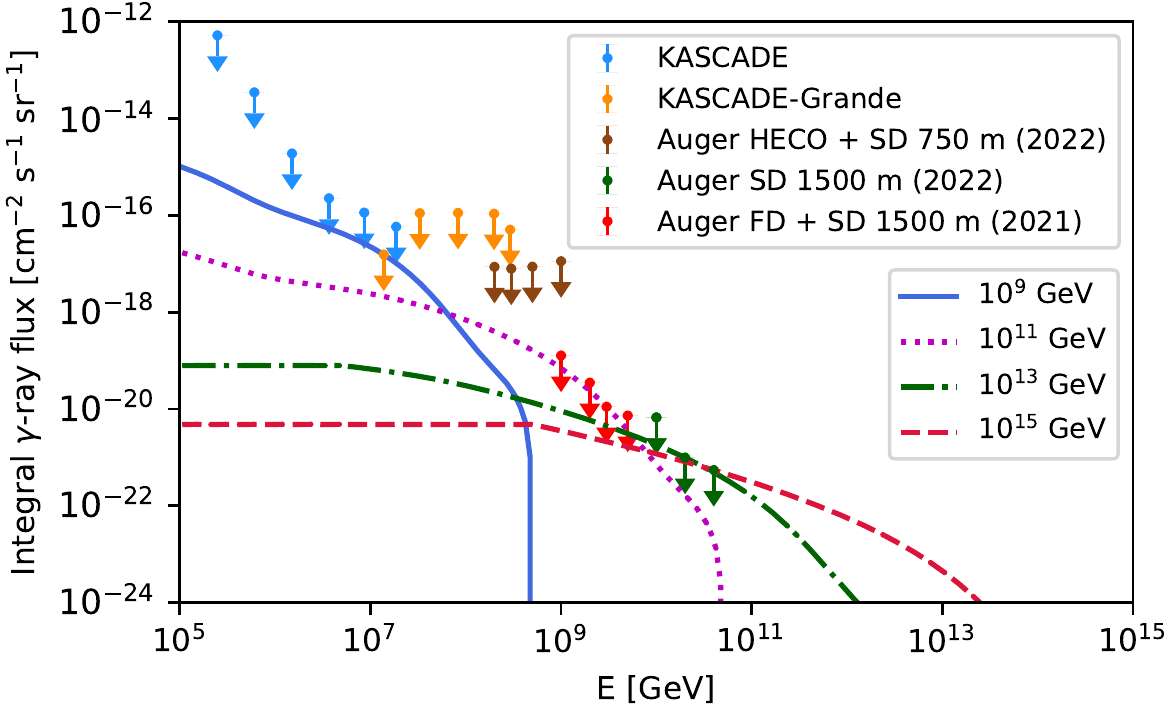}
\caption{\small{\textit{Left:} $p+\overline{p}$ fluxes at Earth from Galactic + extragalactic DM combined for different values of DM mass $m_\chi=10^{9}$, $10^{11}$, $10^{13}$, and $10^{15}$ GeV; decaying through the $b\overline{b}$ channel. The black-colored data points are UHECR spectrum data from \cite{PierreAuger:2019phh}. The brown-colored upper limits at the highest energy bins are derived using the surface detector data. An extrapolation of the upper limit at higher energies is shown by the dotted line. \textit{Right:} Integrated $\gamma$-ray fluxes at Earth from the Galactic DM for discrete values of DM mass $m_\chi=10^{9}$, $10^{11}$, $10^{13}$, and $10^{15}$ GeV; decaying through $b\overline{b}$ channel. The upper limits on the flux from KASCADE, KASCADE-Grande \citep{KASCADEGrande:2017vwf}, and Pierre Auger Observatory \citep{PierreAuger:2022uwd, Savina:2021cva, PierreAuger:2022aty} are shown. The KASCADE limits are converted from 90\% C.L. to 95\% C.L. assuming Poisson statistics. We consider the NFW density profile for DM distribution.}}
\label{fig:dm}
\end{figure*}

\begin{figure*}
\centering
\includegraphics[width=0.49\textwidth]{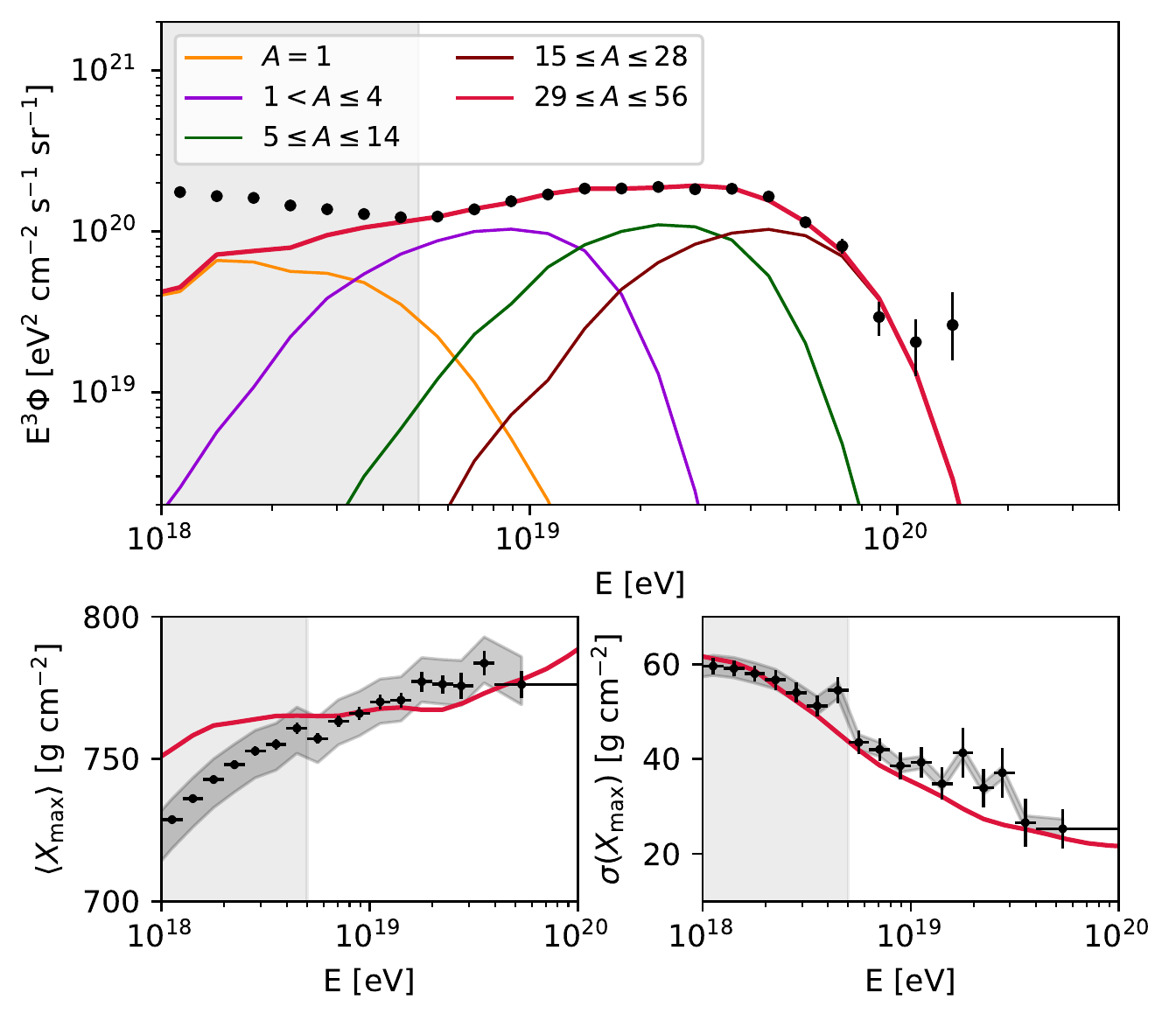}%
\includegraphics[width=0.49\textwidth]{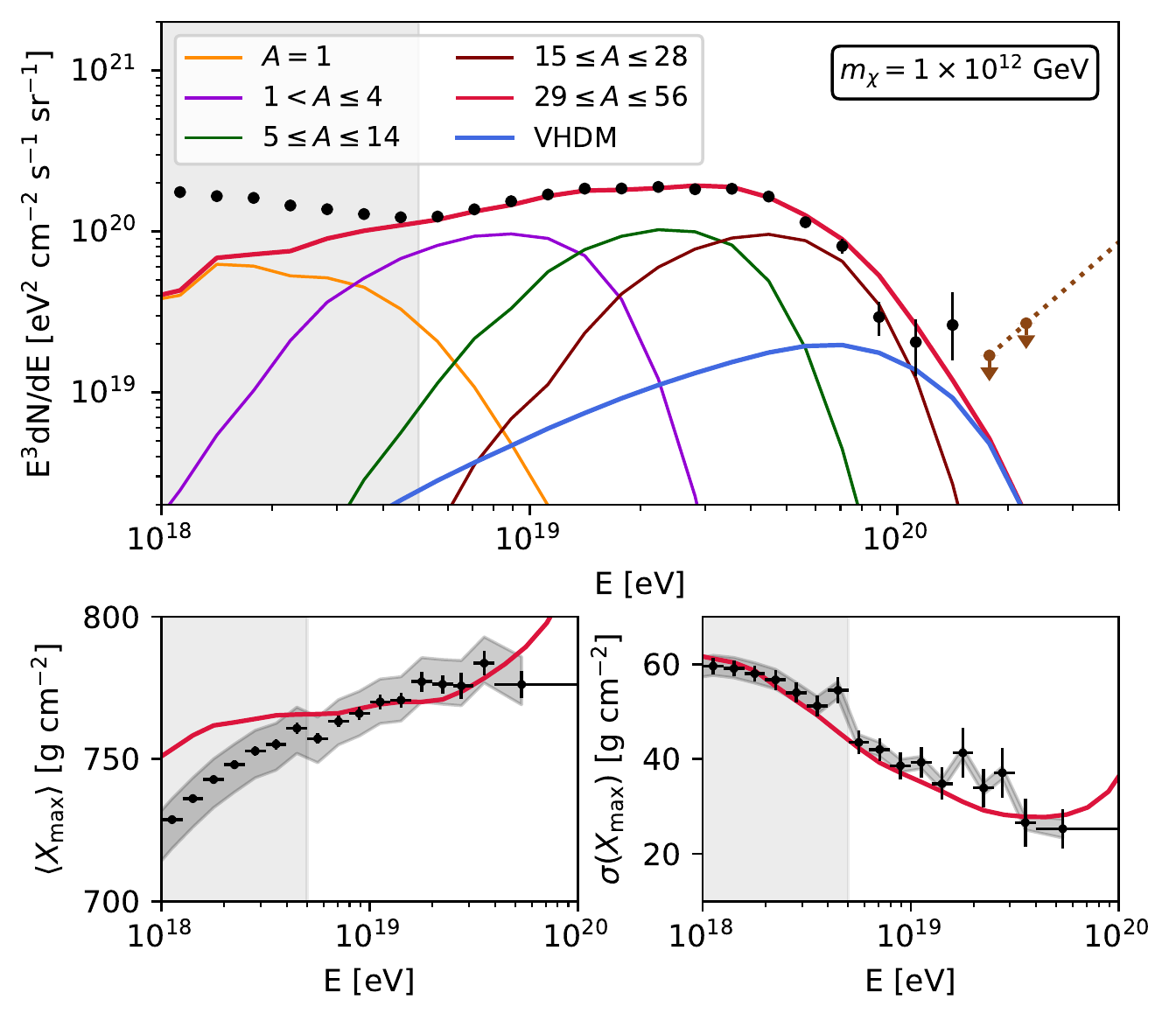}
\caption{\small{The left panel shows the simulated UHECR spectrum, $X_{\rm max}$, and $\sigma(X_{\rm max})$ for the best-fit source parameters obtained by a combined fit of spectrum and composition with the Auger data. The shaded region is excluded from the fit. Here only astrophysical contribution is assumed from a homogeneous source distribution. The right panel shows the simultaneous contribution from the astrophysical and DM components for $m_\chi=1\cdot10^{12}$ GeV}. The fractional contribution from VHDM decay corresponds to 95\% C.L. value of $\chi^2$ statistic. See text for more details.}
\label{fig:astro+dm}
\end{figure*}



\begin{figure*}
\vspace*{0.2cm}
\includegraphics[width=.98\textwidth]{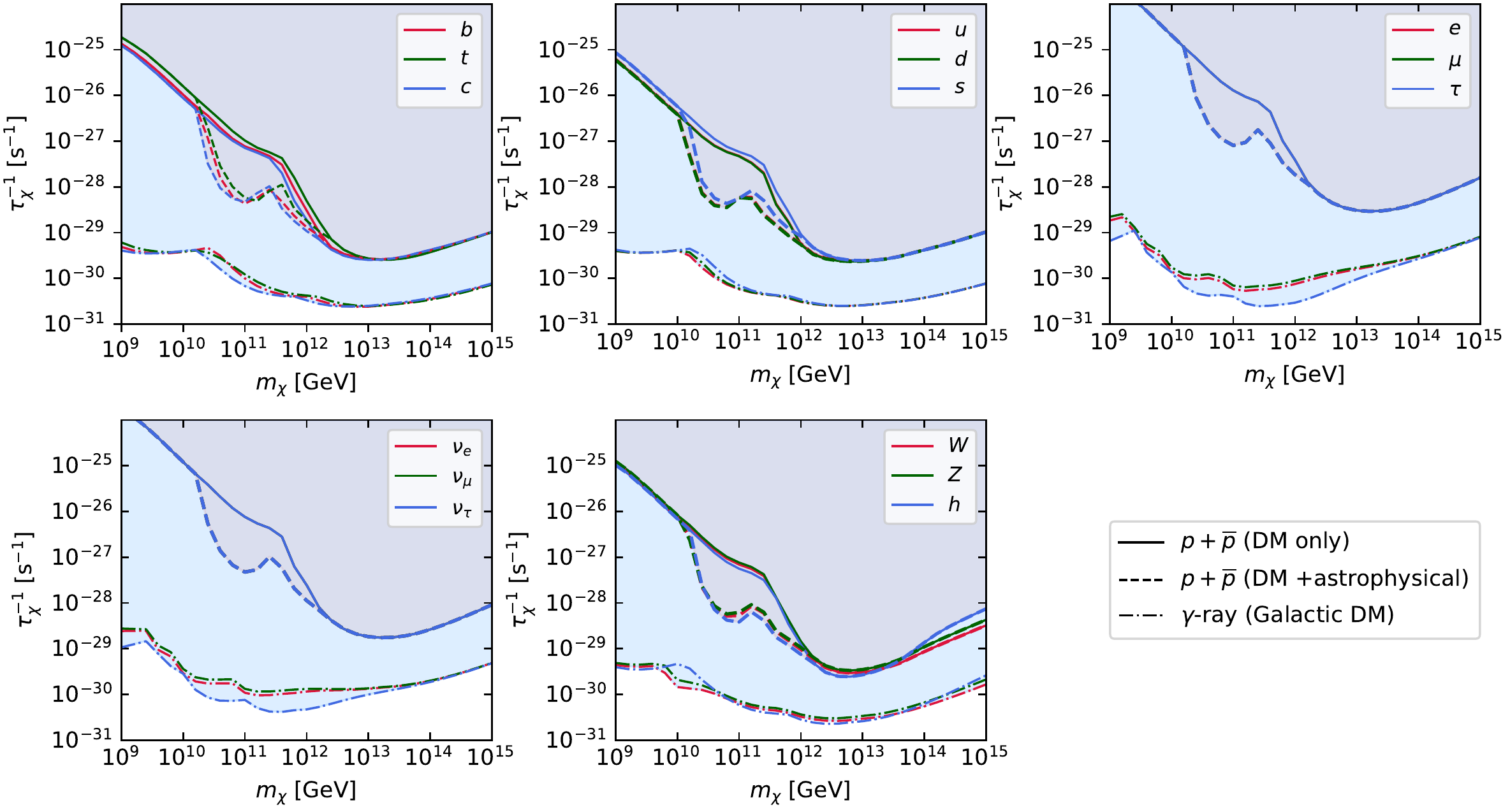}
\caption{\small{DM decay rate constrained by the observed UHECR flux. The solid line corresponds to $p+\overline{p}$ constraints from only the DM scenario and the dashed line is a more stringent limit when both DM (Galactic + extragalactic) and astrophysical components are considered. A number of initial states from prompt DM decay are considered as indicated in the plot labels. It can be seen that the results obtained for all the leptons $e$, $\mu$, and $\tau$ are identical. The same is applicable to different neutrino flavors and vector bosons. The dashed-dotted lines indicate $\gamma$-ray constraints from Galactic DM decay.}}
\label{fig:tau_dm}
\end{figure*}

The cosmic-ray flux in the energy range of study is dominated by the Galactic component, as seen in Fig.~\ref{fig:dm}. Hence, we can write,
\begin{align}
E_p^2\Phi_p &\approx \dfrac{R_{\rm sc} J_\Omega}{4\pi}\dfrac{\rho_{\rm sc} c^2} {\tau_\chi \mathcal{R}_p} \simeq 3.6\times 10^{-11}\times\bigg(\dfrac{\mathcal{J}_\Omega}{2}\bigg) \nonumber \\ 
& \times \bigg(\dfrac{\mathcal{R}_p}{50}\bigg)^{-1} \bigg(\dfrac{\tau_{\chi}}{10^{30} \rm{\ s}}\bigg)^{-1} \ \rm{GeV cm}^{-2}\rm{s}^{-1}sr^{-1}
\end{align}
where $\tau_{\chi,29}= \tau_\chi/10^{29}$~s and $E^2dN_s/dE \approx m_\chi/\mathcal{R}$. Using the publicly available code \textsc{HDMSpectra} \citep{Bauer:2020jay}, we find that for the $b\overline{b}$ channel, the value of $\mathcal{R}$ is roughly found to be $\approx 45$ for decay into protons and $\approx 17$ for decay into $\gamma$-rays. Thus, for $\gamma$-rays, we can write the differential flux as
\begin{align}
E_\gamma^2\Phi_\gamma & \simeq 0.9\times 10^{-10} \times\bigg(\dfrac{\mathcal{J}_\Omega}{2}\bigg) \nonumber \\ 
& \times \bigg(\dfrac{\mathcal{R}_\gamma}{20}\bigg)^{-1} \bigg(\dfrac{\tau_{\chi}}{10^{30} \rm{\ s}}\bigg)^{-1} \ \rm{GeV cm}^{-2}\rm{s}^{-1}sr^{-1}
\end{align}

\subsection{UHECRs \protect}
The hybrid data of cosmic-ray flux measured by Auger, i.e., using both surface and fluorescence detectors, extend up to $\log_{10}(E/\rm eV)= 20.15$. An analysis considering 100\% efficiency of the surface detector above $10^{20}$ eV can impose 90\% C.L. upper limits on the UHECR flux up to $\log_{10}(E/\rm eV)= 20.35$ \citep{PierreAuger:2020qqz, PierreAuger:2020kuy}. A linear extrapolation of the upper limits in the logarithmic energy scale serves as a constraint for UHECR flux from VHDM decay at these extreme energies. This, in turn, provides a lower bound to the DM decay timescale $\tau_\chi$. The extragalactic $p$ and $\overline{p}$ lose energy via interactions with the cosmic background photons, viz., the cosmic microwave background (CMB) and the extragalactic background light (EBL) consisting of IR/UV/optical photons. We use the Gilmore et al. EBL model throughout our analysis \cite{Gilmore:2011ks}.

The flux of the extragalactic component is orders of magnitude lower than the Galactic component due to higher energy losses. The lower limit to $\tau_\chi$ is found by the condition that the simulated flux in any energy bin $i$ is $J_i\leqslant M_i + n\times \Sigma_i$ where $M_i$ is the observed cosmic-ray flux and $\Sigma_i$ is the error in the $i$-th energy bin \cite{Fermi-LAT:2010qeq}. The values of $n=1.28$, 1.64, 4.3 corresponds to 90, 95, and 99.9999\% C.L. lower limits. The observed $p+\overline{p}$ flux at Earth from DM decay in $\chi\rightarrow b \overline{b}$ channel is shown in the left panel of Fig.~\ref{fig:dm} for discrete values of $m_{\chi}$ in the range $10^{9}$ GeV $\leq m_\chi \leq 10^{15}$ GeV. The fluxes correspond to 95\% C.L. lower limit on $\tau_\chi$. It can be seen that for $m_\chi\gtrsim 10^{13}$ GeV, the flux upper limits from Auger constrain the low energy tail of the DM decay spectrum and plays a crucial role in providing improved constraints.

Next, we calculate the values of $\tau_\chi$ when both astrophysical and DM components are present. For this, we first determine the best-fit astrophysical model to Auger spectrum and composition data. The best-fit values of the UHECR source parameters are obtained by scanning over the grid of plausible ranges. We consider a uniform source distribution over comoving distance at 1-5000 Mpc, injecting cosmic-ray nuclei with energy between 1-1000 EeV, according to the spectrum
\begin{align}
\dfrac{dN_A}{dE} =  f_A J_0 \left(\dfrac{E}{E_0}\right)^{-\alpha} \times
\begin{cases}
1\\
\exp\bigg(1-\dfrac{E}{Z_AR_{\rm cut}}\bigg)
\end{cases}
\end{align}

The cutoff rigidity $\log_{10}(R_{\rm cut}/V)$ is varied in the range $[18.0, 18.6]$ at intervals of 0.1, and the spectral index $\alpha$ is varied with grid spacing of 0.1. We assume a mixed composition of representative stable nuclei Hydrogen ($^1$H), Helium ($^4$He), Nitrogen ($^{14}$N), Silicon ($^{28}$Si), and Iron ($^{56}$Fe). We use the Gumbel distribution function $g(X_{\rm max}| E, A)$ to calculate the distribution of the maximum of shower depth $\langle X_{\rm max}\rangle$ and the shower-to-shower fluctuation $\sigma(X_{\rm max})$ \cite{Domenico_2013}. The increasing mass of primary particles at higher energies suggests that the spectral hardening near the ankle cannot arise from the Bethe-Heitler pair-production losses of primary protons \cite{PierreAuger:2016use}, and thus suggests distinct source populations would be required for explaining the spectrum below and above the ankle. Hence we only fit the part of the data for $E>5\times10^{18}$ eV, i.e., beyond the ankle. The goodness-of-fit is calculated using the $\chi^2$ statistic,
\begin{align}
\chi^2_j = \sum_i \dfrac{[J_i^{\rm sim}(E'_i; f_i)-J_i^{\rm ob}(E_i)]^2}{\sigma_i^2} + \bigg(\dfrac{\delta_E}{\sigma_E}\bigg)^2
\end{align}
where the summation runs over all energy bins $i$ included in the fitting procedure and $j$ corresponds to each of the three observables, viz., the energy spectrum, $X_{\rm max}$, and $\sigma(X_{\rm max})$. The systematic error in the Auger spectrum data is dominated by the 14\% energy uncertainty $\sigma_E$. We introduce a nuisance parameter $\delta_E$ such that $E'_i = (1+\delta_E)E$, where $\delta_E$ is varied in the range $-0.14\leqslant\delta_E\leqslant+0.14$ to find the lowest $\chi^2$. The flux normalization of the simulated spectrum is also treated as a free parameter. The differential energy budget required can be expressed as $EQ_E = E^2dN/dE$. The resulting best-fit case is shown in the left panel of Fig.~\ref{fig:astro+dm}. The corresponding parameter values are $\log_{10}(R_{\rm cut}/{\rm V})=18.2$, $\alpha=-1.2$, $f_{\rm H}=0.68$, $f_{\rm He}=0.31$, $f_{\rm N}=0.01$, $f_{\rm Si}=0.0005$, and $f_{\rm Fe}=0$. The corresponding $\chi^2$/d.o.f = 21.03/26. The local (z=0) energy rate density for protons is $\approx 7\times10^{44}$ erg Mpc$^{-3}$ yr$^{-1}$. A detailed analysis of the combined fit of the energy spectrum, mass composition, and arrival direction of the Auger data suggests a redshift evolution of the form $(1+z)^m$ with the best-fit power-law index $m=3.4$ \citep{PierreAuger:2021oxo}. However, for simplicity, we have considered a homogeneous source distribution over redshift.

Now, we vary the normalization of the astrophysical component, keeping all other parameters fixed, to add $p+\overline{p}$ fluxes from DM decay, so that $\Phi = A_1 \Phi_\chi + A_2 \Phi_{\rm astro}$. The 95\% C.L. lower limit to $\tau_\chi$, in this case, is obtained from the value of $A_2$ that gives $p$-value = 0.0455 (32 d.o.f) for the combined \km{$\chi^2$} fit at $E\gtrsim10^{18.7}$ eV. A representative case is shown in the right panel of Fig.~\ref{fig:astro+dm} for $b\overline{b}$ decay mode and $m_\chi=10^{12}$ GeV. In some cases, an improvement in the composition fit can be obtained with the addition of a proton component of DM origin. This is also predicted in earlier studies using two source populations to fit the highest energy spectrum, and composition of UHECRs \cite{Muzio:2019leu, Das:2018ymz}. Here we do not quantify the significance of such a two-population model by variation of source parameters but can be found in earlier works \citep{Das:2018ymz}, which suggests the proton fraction at the highest energy bin can be $10\%-15\%$ at $3.5\sigma$ statistical significance. However, for $m_\chi>10^{11}$ GeV, the astrophysical flux is negligible, and hence DM component becomes dominant, restricted by the flux upper limits at the low energy tail of $J_\chi$. The procedure is repeated for other DM decay channels.

\subsection{UHE photons}
The contribution to $\gamma$-ray fluxes from extragalactic dark matter is negligible in our energy range of interest. Also, the mean free path of $\gamma$-rays from the prompt DM decay is larger than the Galactic length scales, and hence the cascades can be neglected for the Galactic contribution. At lower energies $E_\gamma<10^{9}$ GeV, we use the isotropic diffuse $\gamma$-ray flux upper limits from KASCADE and KASCADE-Grande \cite{KASCADEGrande:2017vwf} to constrain $\tau_\chi$ for $ 10^{9}$ GeV $\lesssim m_\chi \lesssim 10^{12}$ GeV. A conversion factor was taken into account, assuming no background \km{with Poisson statistics}, to obtain 95\% C.L. upper limits. At higher energies, we use the latest Auger SD upper limits for the first time in this work \cite{Savina:2021cva, PierreAuger:2022uwd, PierreAuger:2022aty}, which gives the best up-to-date constraints on the VHDM lifetime at $m_\chi\gtrsim 10^{12}$ GeV. For $\gamma$-rays below a few times $10^{10}$ GeV, the attenuation due to $\gamma\gamma$ absorption in CMB is appreciable. We use an attenuation factor of $\exp(-\tau^{\rm CMB}_{\gamma\gamma})$ in Eqn.~\ref{eqn:los} obtained using the parametrization given in Ref.~\citep{Vernetto:2016alq}. The effects of infrared and optical photons in the interstellar radiation field can be neglected for our energy range of interest.

In the right panel of Fig.~\ref{fig:dm}, we show the integrated $\gamma$-ray fluxes from DM decay, constrained by the integrated $\gamma$-ray limits, using the same calculations as in Eqn.~\ref{eqn:los}--\ref{eqn:phi_gal}. The integrated flux saturates below a specific energy due to the cutoff in the prompt DM decay spectrum at $2E/m_\chi\sim 10^{-6}$, obtained using \textsc{HDMSpectra}. Improvement in the Auger SD limits by $\gtrsim40\%$ provides tighter bounds than those obtained in earlier studies. Fig.~\ref{fig:tau_dm} shows the $\tau_\chi^{-1}$ as a function of $m_\chi$ as obtained in this analysis for various channels of DM decay and both cosmic rays and $\gamma$-rays. It can be seen the limits imposed by $\gamma$-ray constraints are more stringent than that from cosmic rays. The gray-shaded region corresponds to that excluded by both cosmic-ray and $\gamma$-ray constraints, while the white region is the allowed range at 95\% C.L.

%
%

The cosmic-ray flux constrains $\tau_\chi$ to $\gtrsim 4\times 10^{29}$~s at $10^{13}$ GeV for the $q\overline{q}$ decay channel. However, we find that the $\gamma$-ray flux upper limits constrain the VHDM lifetime to $\tau_\chi\gtrsim 4\times 10^{30}$~s at $10^{13}$ GeV for the $b\overline{b}$ channel, which is an order of magnitude longer than the former. This implies the local DM energy budget to be $\lesssim 4.4\times 10^{41}$ erg Mpc$^{-3}$ yr$^{-1}$. The Auger cosmic-ray data bounds the DM lifetime to $\tau_\chi\gtrsim 10^{29}$~s for $10^{12}\lesssim m_\chi \lesssim 10^{15}$ GeV in the $b\overline{b}$ decay mode. 


\section{\label{sec:discussions}Discussion\protect}
We took into account the cosmic-ray flux originating from extragalactic astrophysical sources for a model-dependent estimate of the DM lifetime at $\approx$95\% C.L. The astrophysical component, in our study, is obtained by a combined fit of the spectrum and composition measurements by Auger. Adding $p+\overline{p}$ flux from DM decay leads to the improvement in the combined fit in some cases. The resulting value of $\tau_\chi$ varies by a maximum of one order of magnitude in the energy range between $1.5\times 10^{10}$ GeV and $1.5\times 10^{12}$ GeV, in comparison to the only DM scenario. The most stringent upper limits are obtained for the quark and boson decay channels as shown in Fig.~\ref{fig:tau_dm}, where we obtained $\tau_\chi \gtrsim 10^{29}$~s for $10^{12}\leq m_\chi\leq 10^{14}$ GeV. Our results from cosmic-ray fluxes restricted by the Auger data are more stringent, due to the incorporation of astrophysical fluxes, than the constraints obtained in earlier studies \cite{Ishiwata:2019aet}. The latter found $\tau_\chi\gtrsim$ a few times $10^{28}$~s for $10^{12}$ GeV $<m_\chi<10^{14}$ GeV, bound from the Auger cosmic-ray data. However, compared to the latter, the constraints from cosmic rays are improved due to the addition of the astrophysical fluxes. In addition, we take into account the UHECR composition data ($\langle X_{\rm max}\rangle$ and $\sigma(X_{\rm max})$), as well as explore a wide range of lepton, quark, and gauge boson decay modes. The composition data prefers heavier nuclei, with progressively increasing energy up to the highest energy bin of $\approx 3.5\times10^{19}$ eV, and thus allows a limited contribution from DM decay fluxes in the corresponding energy range. 
We note that such long lifetimes in the explored mass range are also constraining the structure of the interactions in the dark sector, as recently shown in Refs.~\citep{PierreAugerCollaboration:2022tlw, PierreAugerCollaboration:2022wir}.

The Auger field of view is restricted over the angular range $0\leq \alpha_f \leq 2\pi$ in right ascension and the declination band $-\pi/2 \leq \delta_f \leq +\pi/4$. In principle, it is possible to calculate the solid angle averaged $\mathcal{J}$ factor (given in Eqn.~\ref{eqn:phi_gal}) over the Auger field-of-view. Using the NFW model, this results in $\approx 5\%$ change in the $\tau_\chi$ estimates, deduced from Galactic $\gamma$-ray flux. Again, the uncertainties in the DM profile can lead to uncertainty in the sensitivity of detectors \citep{Guepin_2021}. We find that using the Einasto density profile, the resulting change is less than 5\% of the values obtained using the NFW profile. 

\begin{figure}
    \centering
    \includegraphics[width=0.48\textwidth]{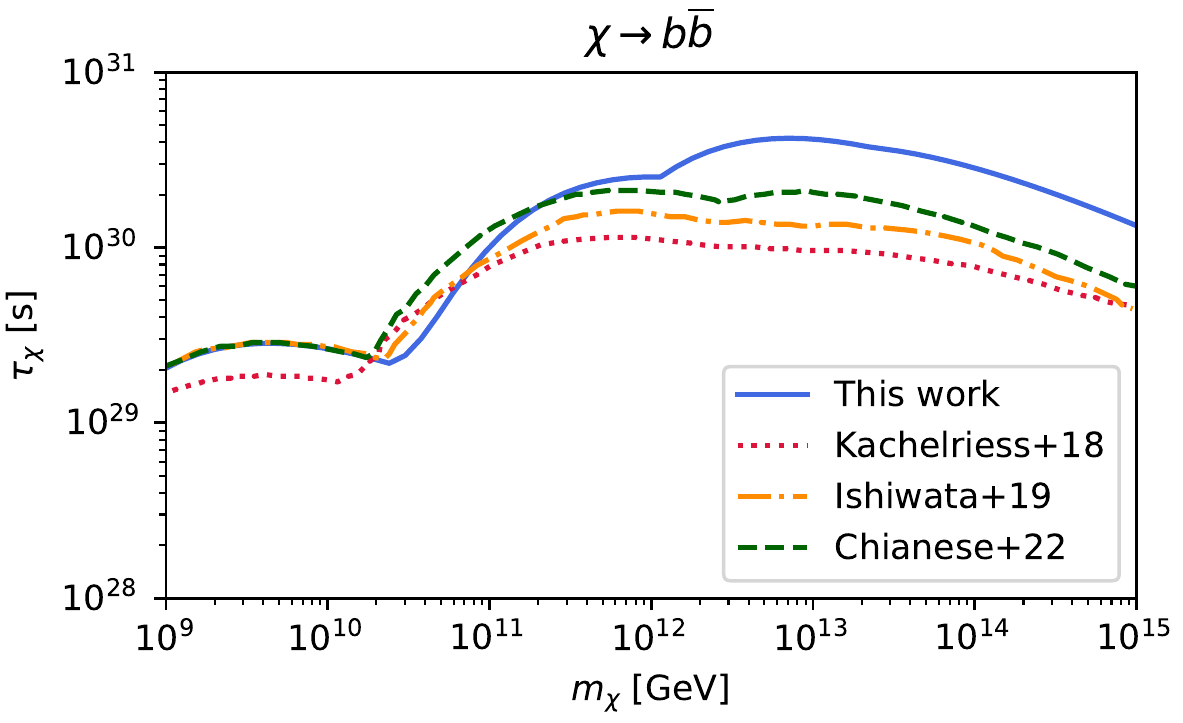}
    \caption{\small{Comparison of the 95\% CL lower limit to VHDM decay timescale through $b\overline{b}$ channel, with existing $\gamma$-ray limits \citep{Kachelriess:2018rty, Ishiwata:2019aet, Chianese:2021jke} in the mass range $10^{9}-10^{15}$ GeV.}}
    \label{fig:compare}
\end{figure}

Heavy DM ($\gtrsim 1$ TeV) beyond the electroweak scale can be produced nonthermally, and it may decay or annihilate in the late universe, and high-energy $\gamma$ rays are cascaded down to GeV-TeV energies. 
Current $\gamma$-ray experiments are sensitive in the GeV-TeV energy band. 
Imaging atmospheric Cherenkov telescopes, e.g., H.E.S.S., VERITAS, and MAGIC, as well as other air-shower detectors such as HAWC and LHAASO are crucial to probe DM signals from the Galactic center direction \citep{HESS:2016mib, HAWC:2017udy, MAGIC:2016xys}. 
A better understanding of the Galactic diffuse emission is crucial for the indirect measurement, especially for the annihilating DM signal. 
In this work, we focus on decaying DM with $10^{9}$ GeV $\leq m_\chi \leq$ $10^{15}$ GeV using Galactic and cosmological DM decay, constrained by the latest cosmic-ray data and $\gamma$-ray flux upper limits obtained from Auger.

The integrated $\gamma$-ray flux upper limits from Auger provide tighter constraints on the DM decay lifetime. Among different final states, we found $\chi\rightarrow b\overline{b}$ and $\chi\rightarrow hh$ give the most severe bounds. For the $b\overline{b}$ channel, $\tau_\chi\gtrsim 3\times 10^{30}$~s at $10^{12.2}$ GeV $\lesssim m_\chi \lesssim 10^{13.8}$ GeV. Using the latest Auger SD upper limits for the first time in this work and considering all decay modes into Standard Model particles in our study, our results indicate $\tau_\chi \gtrsim10^{30}$~s for $10^{11}$ GeV $<m_\chi<10^{15}$ GeV. As shown in Fig.~\ref{fig:compare} for the $b\overline{b}$ decay mode, our estimates at $\gtrsim10^{12}$ GeV is improved by a factor of $\sim2$ compared to that obtained in Ref.~\citep{Chianese:2021jke} and by an even higher factor than those obtained in earlier studies \citep{Kachelriess:2018rty, Ishiwata:2019aet}.
Also, in the Fermi-LAT energy range, isotropic diffuse $\gamma$-ray background emission arises from unresolved $\gamma$-ray sources such as blazars and radio galaxies. The Fermi-LAT constraints from the observation of the Galactic halo excluded decay lifetimes $\gtrsim10^{27}$~s for the $b\overline{b}$ decay mode and NFW profile, \citep{Ackermann_2012}, which is improved with cascades \citep{Ando:2015qda, Cohen:2016uyg, Blanco:2018esa}.
We consider only the ultrahigh-energy $\gamma$-rays from prompt DM decay, owing to negligible cascade interactions at Galactic length scales. 

Neutrino constraints, although they are not our focus, are important, especially in the mass range of $10^6$ GeV $\lesssim m_\chi \lesssim 10^{8}$ GeV. Future neutrino and UHECR experiments like POEMMA \citep{POEMMA:2020ykm} and GRAND \citep{GRAND:2018iaj} will significantly improve the sensitivity at higher energies \citep{Chianese:2021htv}. A recent analysis of the expected photon fluxes from UHECR interactions with matter in the Galactic disk puts constraints on the allowed range of DM lifetime for $m_\chi<10^{11}$ GeV \citep{Berat:2022iea}. Our results are consistent with that obtained there using photon fluxes from the Galactic center region. A recent analysis by LHAASO-KM2A reveals no excess in DM signal from the observation of northern $\gamma$-ray sky \citep{LHAASO:2022yxw}. However, they impose strict limits on the lifetime of VHDM particles between $10^5$ and $10^9$ GeV. Our results are complementary to theirs and an extension of the energy range studied by LHAASO. An estimate with further exposure will reveal improved constraints at the ultrahigh-energy range.


\section{\label{sec:conclusions}Summary\protect}
We revisited constraints on VHDM decaying into standard model particles and placed lower limits to the decay timescale at energies higher than $10^{9}$~GeV and extending up to $\approx10^{15}$~GeV, using the latest UHECR data and UHE $\gamma$-ray flux upper limits measured by the Pierre Auger Observatory. At energies beyond $10^{12}$~GeV, the cosmic-ray flux from astrophysical sources is negligible. The upper limits from the Auger surface detector data constrain the flux from DM decay up to $10^{20.35}$~eV. The integrated $\gamma$-ray flux from Auger puts a more stringent constraint on the upper limit to the DM decay rate. We considered two different DM density profiles, viz., the NFW and Einasto profiles to check systematic uncertainties. 
The constraints from Galactic $\gamma$-ray fluxes are an order of magnitude stronger than those obtained using cosmic rays. Observations of ultrahigh-energy $\gamma$ rays with future telescopes will provide a better test of the DM decay lifetime.


\begin{acknowledgments}
We thank Nagisa Hiroshima, Atsushi Naruko, Deheng Song, and Bing T. Zhang for useful discussions related to the work. 
We also thank the Pierre Auger Collaboration and its publication committee for their useful comments and valuable suggestions, especially Olivier Deligny, Carola Dobrigkeit, Ioana Maris, and Marcus Niechciol.
This research of S.D., T.F., and K.M. is supported by KAKENHI No.~20H05852. 
This work of K.M. is supported by the NSF Grant No.~AST-1908689, No.~AST-2108466 and No.~AST-2108467, and KAKENHI No.~20H01901. 
\end{acknowledgments}




\nocite{*}

\bibliography{apssamp}

\end{document}